# Phase Diagram of Infinite Layer Praseodymium Nickelate Pr$_{1-x}$Sr$_x$NiO$_2$ Thin Films

Motoki Osada,[1,2,3,†] Bai Yang Wang,[1,4] Kyuho Lee,[1,4] Danfeng Li,[1,2,5,‡] and Harold Y. Hwang[1,2,§]

[1]*Stanford Institute for Materials and Energy Sciences, SLAC National Accelerator Laboratory, Menlo Park, CA 94025, United States.*

[2]*Department of Applied Physics, Stanford University, Stanford, CA 94305, United States.*

[3]*Department of Materials Science and Engineering, Stanford University, Stanford, CA 94305, United States.*

[4]*Department of Physics, Stanford University, Stanford, CA 94305, United States.*

[5]*Department of Physics, City University of Hong Kong, Kowloon, Hong Kong, People's Republic of China.*

ABSTRACT

We report the phase diagram of infinite layer $Pr_{1-x}Sr_xNiO_2$ thin films synthesized via topotactic reduction from the perovskite precursor phase using $CaH_2$. Based on the electrical transport properties, we find a doping-dependent superconducting dome extending between $x = 0.12$ and 0.28, with a maximum superconducting transition temperature $T_c$ of 14 K at $x = 0.18$, bounded by weakly insulating behavior on both sides. In contrast to the narrower dome observed in $Nd_{1-x}Sr_xNiO_2$, a local $T_c$ suppression near $x = 0.2$ was not observed for the $Pr_{1-x}Sr_xNiO_2$ system. Normal state Hall effect measurements indicate mixed carrier contributions of both electrons and holes, and show a sign change in the Hall coefficient as functions of temperature and $x$, quite similar to that in $Nd_{1-x}Sr_xNiO_2$. Also similar is the observation of a minimum in the normal state resistivity associated with the superconducting compositions. These findings indicate an infinite layer nickelate phase diagram that is relatively insensitive to the rare-earth element, but suggest that disorder arising from the variations of the ionic radii on the rare-earth site affects the superconducting dome.

High-$T_c$ cuprates are unusual materials systems, marked by a rather complex phase diagram enriched with intertwined electronic orders [1,2]. Despite exhibiting a generic doping-dependent phase diagram with common features, which many consider a hallmark of the physics of the $CuO_2$ planes [1], cuprates of various crystal structures [3-7] show rather diverse properties, manifested by different configurations of Cu-O stacking layers [8-10], a remarkably wide range of transition temperatures [1], and varied strength of coupling to proximate competing orders [2,11]. Many of these aspects represent the intimate interplay between the local chemical environment [10], partially set by the cation sublattices outside the Cu-O networks [12-14], and the macroscopic materials properties. Examples include the famous stripe phase [2,15,16] induced by the local instability of the lattice distortion via cation substitution, such as $La_{2-x}Ba_xCuO_4$ [17] and Nd- or Eu-substituted $La_{2-x}Sr_xCuO_4$ [15,18], the latter of which involves a direct tuning of the chemical pressure by varying the rare-earth elements. Such lattice distortion derived rare-earth dependence was also clearly seen in the perovskite nickelates ($ANiO_3$) [19,20], marked by their $A$-site dependent metal-insulator transition and charge-disproportionation temperatures. Furthermore, a rare-earth dependent ground state has been found in the tri-layer nickelates with square-planar coordination, in which the role of the lanthanide elements remains elusive [21].

The recent observation of superconductivity in rare-earth infinite layer nickelates [22,23] has stimulated the discussion on the relationship to their cuprate counterparts, in that they share the same atomic structure and starting $3d^9$ electronic configuration, yet perhaps hosting distinct electronic structures and low-energy physics with a multi-orbital nature, as noted by Lee and Pickett [24]. More recent theoretical considerations [25-31] and scattering/spectroscopic experiments [32,33] highlight the unusual role of rare-earth orbitals and their interaction and hybridization. This raises the question of whether the nickelate superconductors are close cousins to the heavy fermion superconductors [34-36] or intermetallic superconductors, such as $RENi_2B_2C$ ($RE$ = rare-earth) [37]. The latter family of compounds shows a strong dependence of magnetic ordering and superconductivity on the lanthanide element [38], in which the two phenomena seem to correlate. However, superconductivity with similar $T_c$ values observed in thin film $Nd_{0.8}Sr_{0.2}NiO_2$ [22] and $Pr_{0.8}Sr_{0.2}NiO_2$ [23] suggests that the emergence of superconductivity seems relatively insensitive to the details of the rare-earth 4$f$ electron configuration [39-41]. In this regard, the question of whether these doped infinite layer nickelate superconductors host mixed aspects [40,42-45] of both cuprates and superconductors with hybridized rare-earth contributions calls for further investigation.

With this notion, and motivated by the observation of a doping dependent superconducting dome in thin film Nd$_{1-x}$Sr$_x$NiO$_2$ [46, 47], we have investigated the phase diagram across a series of Pr$_{1-x}$Sr$_x$NiO$_2$ (0 ≤ $x$ ≤ 0.32) thin films. We report a superconducting dome structure in the phase diagram of the Pr$_{1-x}$Sr$_x$NiO$_2$ family with a maximum $T_c$ of 14 K at $x$ = 0.18. We show that superconductivity is bound by $x$ = 0.12 and 0.28, beyond which the system exhibits weakly insulating behavior at low temperatures at both ends of the superconducting dome. While the behavior across the different phase boundaries is in qualitative resemblance to that in the Nd series, the fine structure of the superconducting dome shows rather different features, namely the absence of the $T_c$ suppression observed in Nd$_{1-x}$Sr$_x$NiO$_2$ and a broader superconducting doping range. Nevertheless, the Hall effect data in both systems show quantitatively very similar trends, marked by continuous zero crossings of the Hall coefficient ($R_H$) in temperature and doping. These data indicate the band configuration and Fermi surface structure are likely somewhat insensitive to the rare-earth site, consistent with density functional theory (DFT) calculations [39,40].

Pr$_{1-x}$Sr$_x$NiO$_2$ films of 6.3 - 9.7 nm in thickness were grown by pulsed laser deposition, followed by a subsequent topotactic reduction reaction, using the same conditions as described previously [23]. In that study, we found that an upper SrTiO$_3$ (STO) capping layer was not essential for stabilizing a uniform single-crystalline infinite layer structure in PrNiO$_2$ and Pr$_{0.8}$Sr$_{0.2}$NiO$_2$ thin films, which can be possibly attributed to a larger tolerance factor and a better lattice matching to the STO substrate. Therefore, the $Pr_{1-x}Sr_xNiO_3$ films were synthesized without a $SrTiO_3$ capping layer. The substrate temperature was kept at 570 °C. The laser fluence and oxygen pressure for the growth of undoped $PrNiO_3$ and doped $Pr_{1-x}Sr_xNiO_3$ ($x \neq 0$) are 1.39 J/cm², 200 mTorr and 2.19 J/cm², 250 mTorr, respectively. The laser repetition was 4 Hz for all samples. The x-ray diffraction (XRD) characterization and transport measurements were performed as described previously [23].

High-quality precursor $Pr_{1-x}Sr_xNiO_3$ films were synthesized, which facilitates a successful topotactic transition to the resultant infinite layer $Pr_{1-x}Sr_xNiO_2$. In this way, the wide sample-to-sample variation in the temperature dependent resistivity, $\rho_{xx}(T)$, pertaining to $Nd_{1-x}Sr_xNiO_2$ [22,48], is quite a bit narrowed here. Based on this, for simplicity, throughout this report we show one dataset of representative samples for each doping level. Figure 1 illustrates the structural properties of the representative $Pr_{1-x}Sr_xNiO_2$ thin films for each $x$, characterized by XRD measurements. Across all $x$, we observe clear film peaks in the symmetric $\theta$-$2\theta$ scans [Fig. 1(a)] corresponding to the infinite layer phase, which has reduced lattice parameters along the $c$-axis, as compared to the perovskite precursor phase [Fig. 1(b)]. The decreasing trend of the 001 peak intensity with Sr substitution is consistent with our calculation of the powder diffraction pattern in $Pr_{1-x}Sr_xNiO_2$ (not shown), where the ratio of 001 peak intensity to 002 peak intensity is 0.44 at $x = 0$ and monotonically decreases to 0.22 at $x = 0.32$. This structure factor variation due to Sr substitution should be accounted for when considering the overall lower 001 peak intensity upon increasing $x$. Although varied in intensity, the positions of the film peaks clearly shift to a lower angle as $x$ increases, consistent with a monotonic increase of the $c$-axis lattice constants in the range from 3.31 Å ($x = 0$) to 3.44 Å ($x = 0.32$) [Fig. 1(c)]. Reciprocal space maps (RSM) around the $\bar{1}03$ diffraction peak [Fig. 1(d)] confirm that the films are in all cases strained to the substrate. Consequently, the $a$-axis lattice constants match that for STO, which is 3.91 Å [Fig. 1(c)]. The calculated unit-cell volume of PrNiO$_2$ is slightly larger than that of the thin film NdNiO$_2$ on STO [46], the value of which is indicated as a reference in Fig. 1(c). These are in line with the fact that the ionic radius of Pr$^{3+}$ (113 pm) is larger than Nd$^{3+}$ (111 pm) in the same crystal environment and closer to that of Sr$^{2+}$ (126 pm) [49].

The temperature dependent resistivity $\rho_{xx}(T)$ for thin film Pr$_{1-x}$Sr$_x$NiO$_2$ down to 2 K is presented in Fig. 2(a) and 2(b). For both the under- and over-doped regime ($x$ = 0, 0.04, 0.28 and 0.32), $\rho_{xx}$ shows a minimum around a temperature, $T_{min}$, of 70 K, 45 K, 25 K and 40 K, respectively, below which an upturn appears. Despite having a smaller $T_{min}$ with respect to the undoped case, the overall resistivity for $x$ = 0.28 and 0.32 is much larger, and their temperature dependence above $T_{min}$ deviates from the remaining curves. This high $\rho_{xx}$, together with the reduced (001) peak intensity seen in the XRD scans [Fig. 1(a)], suggests that both structural and electronic disorder induced by Sr doping may contribute in the over-doped regime. In the intermediate doping levels, i.e. $x$ = 0.16, 0.18, 0.20 and 0.24, all resistivity curves show metallic behavior with a roughly linear temperature dependence approximately above $T_c$, a hallmark of cuprates and organic superconductors [1,11]. At lower temperature, superconducting transitions with zero resistance states are observed [Fig. 2(b)]. One peculiar doping point is $x$ = 0.12, where both a low temperature resistivity upturn and onset of superconductivity are present, demarking the edge of the superconducting dome. From $\rho_{xx}(T)$, we construct a superconducting phase diagram, as shown in Fig. 2(c), where transition temperatures $T_{c,\,90\%R}$ and $T_{c,\,10\%R}$ are defined as the temperature at which the resistivity is 90% and 10% of $\rho_{xx}(T = 20\ K)$, respectively. The doping dependent superconducting dome spans across $0.12 \leq x \leq 0.24$ with the maximum $T_{c,\,90\%R}$ of 14 K at $x$ = 0.18. This dome is somewhat wider than that of the Nd$_{1-x}$Sr$_x$NiO$_2$ system ($0.15 \leq x \leq 0.225$) [46], as shown in the inset of Fig. 2(c).

Furthermore, the dome (or the functional form of $T_c(x)$) is more of a "bell" shape. Around the optimal doping level ($x = 0.18$), $T_c$ is higher and the transition width (difference between $T_{c, 90\%R}$ and $T_{c, 10\%R}$) is arguably narrower, as compared to that of $Nd_{1-x}Sr_xNiO_2$. All of these aspects likely reflect that $Pr_{1-x}Sr_xNiO_2$ films are generally of higher stability and crystallinity [23]. We also note that the small suppression of $T_c$ observed in the $Nd_{1-x}Sr_xNiO_2$ system is absent here, and discussed further below.

The normal state Hall coefficient $R_H$ ($0.12 \leq x \leq 0.32$) is shown in Fig. 3(a) as a function of temperature. The Hall resistivity $\rho_{yx}$ shows linear magnetic field ($\mu_0 H$) dependence up to 8 T. For all doping levels presented here, $R_H$ increases with decreasing temperature, except for $x = 0.12$, where the $R_H$ maximum associated with $T_{min}$ is present. In particular, the negative $R_H(T)$ for $x = 0.12$ is consistent with the prediction of partially occupied Pr $5d$ derived states [39]. For samples with higher doping ($x \geq 0.20$), $R_H(T)$ crosses zero in a continuous manner, suggesting the mixed carrier contribution of electrons and holes. This multi-orbital nature is also corroborated by the smooth sign change of $R_H$ as function of $x$, for instance at $T = 20$ K, as shown in the top panel of Fig. 3(b), corresponding to an increasing depletion of the electron pockets primarily composed of Pr $5d$ bands [39,40], or a change in the Fermi surface topology upon doping close to half-filling in the strong correlation limit [50].

It is intriguing to see that superconductivity is associated with a small $R_H$ proximate to zero, but of either sign (Fig. 3b; top panel), and the minimum in the normal state $\rho_{xx}$ (Fig. 3b; bottom panel), both of which are plotted in comparison to the similar dataset obtained for $Nd_{1-x}Sr_xNiO_2$ [46]. In particular, the enhanced stability of $Pr_{1-x}Sr_xNiO_2$ allowed us to explore deeper into the heavily doped side of the phase diagram and access the more resistive regime, where both the systematic behavior of $R_H$ and the 'V-shape' normal state $\rho_{xx}(x)$ (with larger scale) persist.

We now discuss the implications of our results. First, our observation of a $T_c$ dome with similar doping dependence and the universal trend of $R_H$ in temperature and doping, in both $Pr_{1-x}Sr_xNiO_2$ and $Nd_{1-x}Sr_xNiO_2$, suggests phenomenologically similar electronic structure [35,36] for both cases. These findings warrant further experimental investigations of the Fermi surface and the band structure, ideally using techniques such as angle-resolved photoemission spectroscopy if experimentally accessible. Next, the absence of the dip in $T_c$ in the superconducting dome of $Pr_{1-x}Sr_xNiO_2$ does not seem to support the existence of a universal 'anomalous' doping level $x$ (0.2 for $Nd_{1-x}Sr_xNiO_2$) across different nickelate materials. Instead, if such a local suppression of superconductivity is considered suggestive of stripe order, as widely seen in cuprates [2,15,17] and other nickelates [51-53], our findings indicates that small differences in strain field or chemical pressure, can affect the ground state of the system. This could be analogous to the different scale of the '1/8' anomaly in $La_{2-x}Sr_xCuO_4$ [54], $La_{2-x}Ba_xCuO_4$ [17], and Nd- or Eu-substituted $La_{2-x}Sr_xCuO_4$ [15,18], and the general role of cation size disorder in determining $T_c$ [13]. While these interpretations are yet to be experimentally established, our results thus far provide evidence that the infinite layer nickelates are susceptible to the delicate balance between the local chemical environment and electronic correlations.

It should be noted that, as we approach the largest values of Sr substitution $x$ in the heavily doped regime, the materials pose increasingly significant challenges to materials synthesis. Firstly, this is because the formal valence of Ni ($Ni^{3+x}$) in the precursor perovskite phase deviates further away from its most thermodynamically stable state ($Ni^{2+}$), which can drive the formation of extended defects to perturb the local Ni valence [48]. In addition, Sr substitution introduces cation-site disorder, as indicated by the systematic reduction of 001 peak intensity as $x$ increases [Fig. 1(a)]. As a consequence, systematic contributions from materials imperfections to $\rho_{xx}$ will likely grow with increasing $x$. In this regard, further careful optimization of growth and investigation of the defect structure, such as the Ruddlesden-Popper-type faults that were observed in $Nd_{1-x}Sr_xNiO_2$ thin films [46,48], in the heavily doped regime is required. However, considering the clear trend in the doping dependent $T_c$ and $R_H$, our finding of a superconducting dome associated with a continuous evolution of the electronic structure with multi-band character should be qualitatively robust.

In summary, we have extended the study of the phase diagram of the infinite layer nickelates, analogous to cuprate systems, by replacing Nd with Pr. Given the greater lattice stability of the precursor perovskite phase $Pr_{1-x}Sr_xNiO_3$ due to the larger tolerance factor and smaller lattice mismatch to STO in the perovskite phase [23], with respect to the previously studied $Nd_{1-x}Sr_xNiO_3$, we were able to stabilize a series of samples to higher doping levels. We observed a $T_c$ dome bound by $x = 0.12$ and $0.28$ in the phase diagram of $Pr_{1-x}Sr_xNiO_2$, beyond which the system exhibits low-temperature weakly insulating behavior at both ends of the doping levels studied here. $R_H$ displays zero-crossing behavior both as a function of temperature and doping, qualitatively in agreement with a multi-band Fermi surface and quantitatively similar to $Nd_{1-x}Sr_xNiO_2$. A continuous variation of $T_c$ with a single maximum of 14 K at $x = 0.18$ and a larger doping range for superconductivity were additionally found, which indicates that cation size disorder effects may play a role in the detailed structure of the superconducting dome.

ACKNOWLEDGMENTS

This work was supported by the US Department of Energy, Office of Basic Energy Sciences, Division of Materials Sciences and Engineering, under contract number DE-AC02-76SF00515, and the Gordon and Betty Moore Foundation's Emergent Phenomena in Quantum Systems Initiative through grant number GBMF9072 (synthesis equipment).

AUTHOR INFORMATION

[†]mosada@stanford.edu

[‡]danfeng.li@cityu.edu.hk

[§]hyhwang@stanford.edu

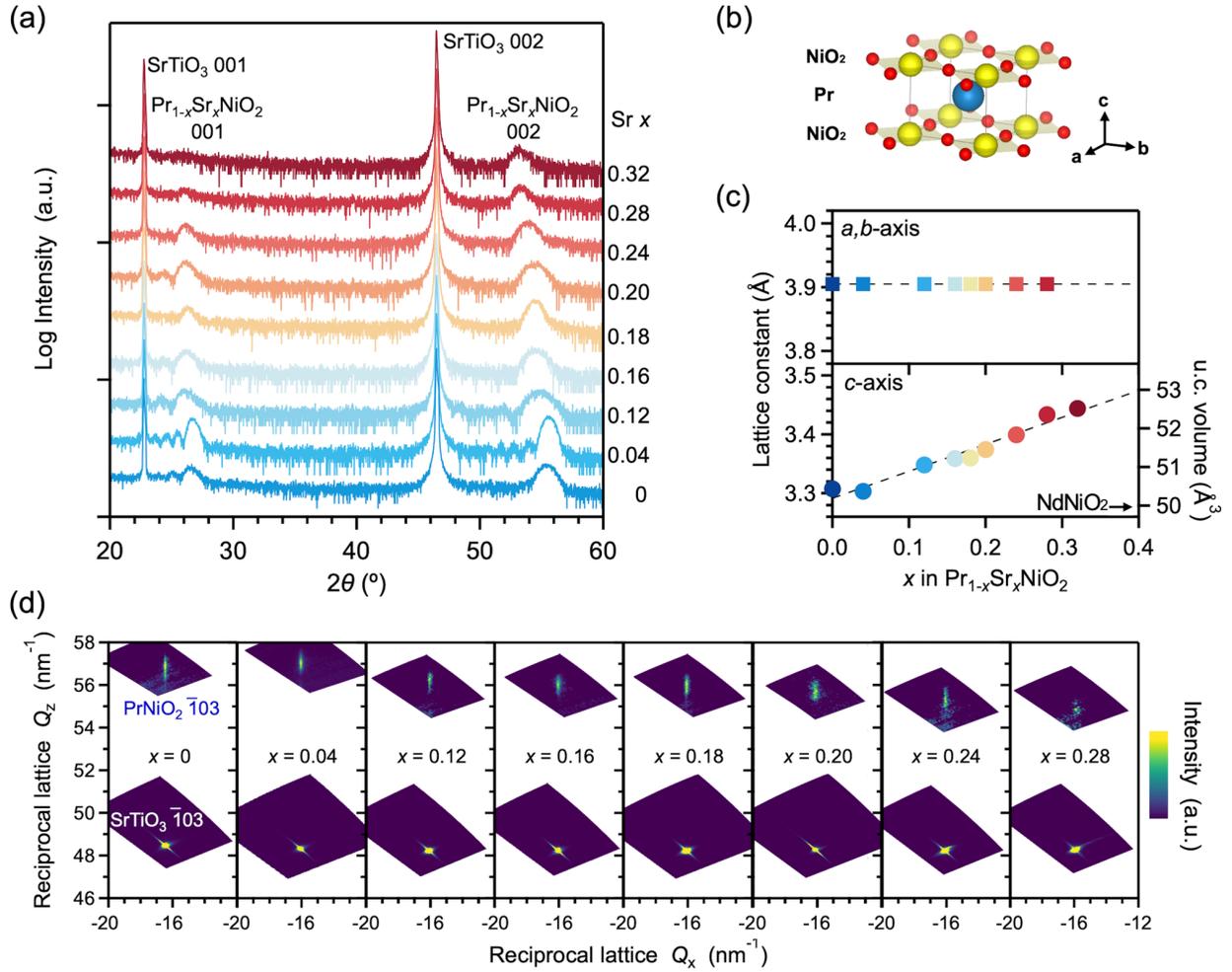

**FIG. 1.** (a) X-ray diffraction $\theta$–$2\theta$ symmetric scans of $Pr_{1-x}Sr_xNiO_2$ thin films grown on $SrTiO_3$ (001) substrates. (b) The atomic structure of the infinite-layer $PrNiO_2$. (c) Lattice constants and unit cell volumes of $Pr_{1-x}Sr_xNiO_2$ thin films. The unit cell volume of thin film $NdNiO_2$ is shown as a reference [46]. (d) Reciprocal space maps around the $\bar{1}03$ peaks. All scans were performed at room temperature.

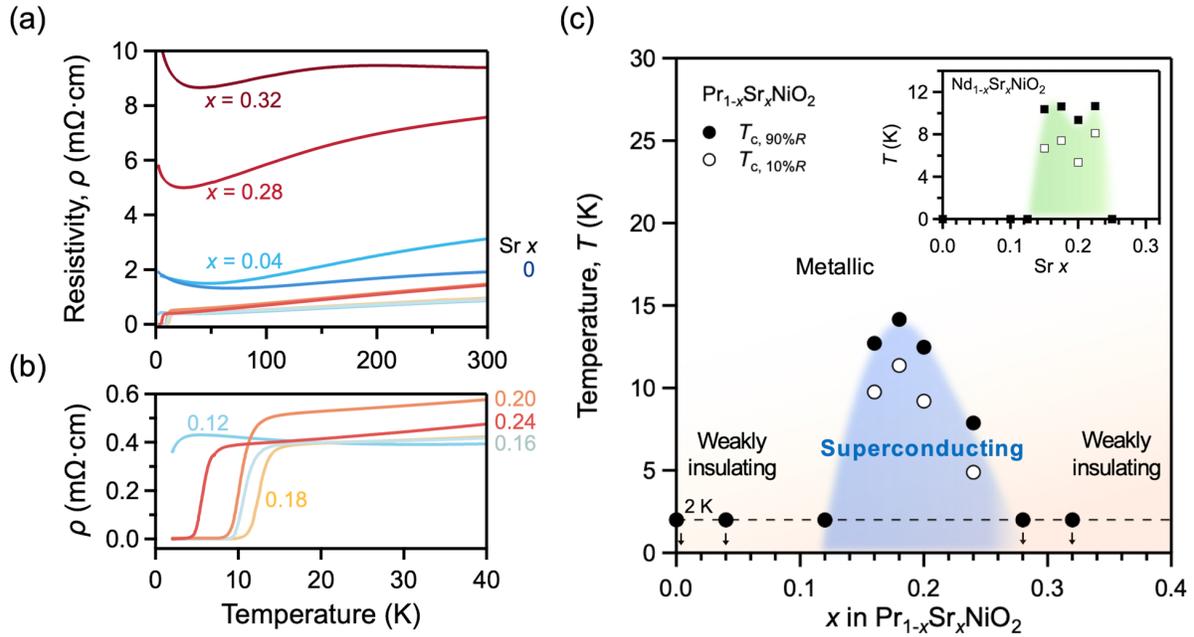

**FIG. 2.** (a) Representative temperature dependence of resistivity $\rho$–$T$ curves for $Pr_{1-x}Sr_xNiO_2$ thin films. (b) The enlarged $\rho$–$T$ curves in the temperature range from 2 K to 40 K. (c) The phase diagram of $Pr_{1-x}Sr_xNiO_2$ thin films. Closed and open circles represent $T_{c,\,90\%R}$ and $T_{c,\,10\%R}$, as defined to be the temperatures at which the resistivity is 90% and 10% of the resistivity value at 20 K, respectively. The dashed line indicates the minimum measurement temperature 2 K. The inset shows the phase diagram of $Nd_{1-x}Sr_xNiO_2$ thin films adapted from [46].

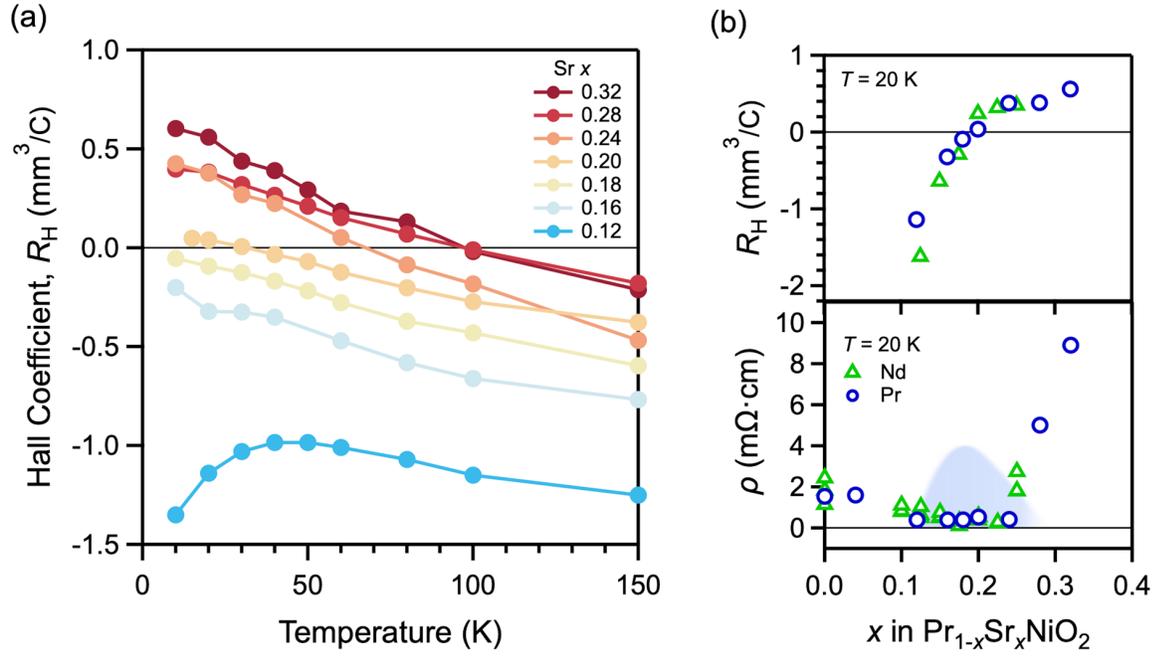

**FIG. 3.** (a) Normal-state Hall coefficient as a function of temperature for $Pr_{1-x}Sr_xNiO_2$ thin films. (b) Hall coefficient $R_H$ and resistivity $\rho$ as a function of Sr $x$ at 20 K (top and bottom, respectively). Triangles and circles represent $Nd_{1-x}Sr_xNiO_2$ [46] and $Pr_{1-x}Sr_xNiO_2$, respectively. The superconducting dome of $Pr_{1-x}Sr_xNiO_2$ thin films is shown in the background of the lower panel.